\documentclass[a4,useAMS,usegraphicx,usenatbib]{mn2e}

\usepackage{graphicx}
\usepackage{color}
\usepackage[colorlinks,bookmarks]{hyperref}
\definecolor{linkblue}{rgb}{0,0,0.8}
\definecolor{linkgreen}{rgb}{0,0.5,0}
\hypersetup{linkcolor=linkblue, citecolor=linkgreen, urlcolor=linkblue}

\usepackage{amsmath}
\usepackage{natbib}

\newcommand{\be}{\begin{equation}}
\newcommand{\ee}{\end{equation}}
\newcommand{\bea}{\begin{eqnarray}}
\newcommand{\eea}{\end{eqnarray}}
\newcommand{\ba}{\begin{eqnarray}}
\newcommand{\ea}{\end{eqnarray}}

\def\d {\mathrm{d}}
\def\S {\mbox{S}}

\newcommand{\aperp}{a_{\perp }}

\newcommand{\apar}{a_{\parallel}}

\newcommand{\Hperp}{H_{\perp}}
\newcommand{\Hperpo}{H_{\perp_0}}
\newcommand{\Hpar}{H_{\parallel}}

\begin{document}

\title[Copernican principle and spatial homogeneity]{Testing the Copernican principle by constraining spatial homogeneity}

\author[W. Valkenburg et al.]{Wessel Valkenburg$^1$, Valerio Marra$^2$ and Chris Clarkson$^3$\\
$^1$Instituut-Lorentz for Theoretical Physics, Universiteit Leiden, Postbus 9506, 2333 CA Leiden, The Netherlands\\
$^2$Institut f\"ur Theoretische Physik, Universit\"at Heidelberg, Philosophenweg
16, 69120 Heidelberg, Germany\\
$^3$Astrophysics, Cosmology and Gravity Centre, and Department of Mathematics and Applied Mathematics, \\\hspace{.2cm}University of Cape Town, Rondebosch 7701, South Africa
}

\date{Accepted XXX. Received XXX; in original form XXX}

\pagerange{\pageref{firstpage}--\pageref{lastpage}} \pubyear{2012}

\maketitle

\label{firstpage}

\begin{abstract}
We present a new programme for placing constraints on radial inhomogeneity in a dark-energy dominated universe. We introduce a new measure to quantify violations of the Copernican principle. 
Any violation of this principle would interfere with our interpretation of any dark-energy evolution.
In particular, we find that current observations place reasonably tight constraints on possible late-time violations of the Copernican principle: the allowed area in the parameter space of amplitude and scale of a spherical inhomogeneity around the observer has to be reduced by a factor of three so as to confirm the Copernican principle.
Then, by marginalizing over possible radial inhomogeneity we provide the first constraints on the cosmological constant which are free of the homogeneity prior prevalent in cosmology.
\end{abstract}


\section{{Introduction}}

The Copernican principle states that humans are not privileged observers of the universe and provides our philosophical basis for assuming that on the largest scales the universe is spatially homogeneous. While it is one of the foundational aspects of modern cosmology, this assumption needs to be tested outside of the standard paradigm~\citep{Bolejko:2008cm,Davis:2010jq,Valkenburg:2011ty,Marra:2012pj}. Though it may seem pedantic to test something so obvious, the standard paradigm itself is built on shaky foundations, relying on an unexplained, gravitationally repulsive, dark-energy component for observations to fit the model. 
As current observations
suggest isotropy around us~\citep{
Blomqvist:2009ps,Komatsu:2010fb},
inhomogeneous models must have 
the observer near the centre of a matter distribution which is approximately spherically symmetric.
Because we can observe changes in the expansion of the universe as a function of redshift only, it remains hard to disentangle temporal evolution  from possible spatial variations spherically around us.
Indeed, there has even been active debate on models whereby
dark energy is replaced by radial inhomogeneity, 
in violation of the Copernican principle (see \citealt{Marra:2011ct,Clarkson:2012bg} for reviews). 
This highlights how intertwined are attempts to detect any evolution of dark energy to the Copernican principle.

To rule out spherically symmetric models, consistency tests have been proposed (see \citealt{Clarkson:2012bg} for a review), but all suffer as they lack at present a quantifiable measure of deviations from homogeneity. 
Attempts have been made to measure the homogeneity of the universe~\citep{Hogg:2004vw,Labini:2010qx,Scrimgeour:2012wt,Jackson:2012jt,Marinoni:2012ba,Keenan:2013mfa}.
Here we introduce a robust approach, by directly constraining any radial change in the density, Hubble rate and curvature assuming that the Copernican principle is \emph{false} and that dark energy exists. We model the universe spherically symmetric, containing both matter and dark energy in the form of a cosmological constant.
We provide marginalized constraints on the amplitude and scale of any spatial inhomogeneity. 
We quantify the deviation from Copernicanism by comparing the observational constraints we find on the spherical inhomogeneity to the theoretical constraints on the existence of the same inhomogeneity in a Copernican universe. 
The ratio of allowed volumes in parameter space describing the inhomogeneity 
gives a measure of how far we are from establishing the Copernican principle. 
Our goal is not to extend the $\Lambda$CDM model,
but to find how `badly' non-Copernican a feature the model can have. 
Occam's razor already favours the Copernican Universe. But Occam's razor 
is a model selection tool. Here, we lay out a programme to \emph{falsify} non-Copernican models, so as to establish the Copernican Principle observationally.

\section{The model}

We model radial inhomogeneity assuming a spherically symmetric Lema\^{i}tre-Tolman-Bondi solution including a cosmological constant $\Lambda$ ($\Lambda$LTB), see e.g.~
\citep{Lemaitre:1933gd,Tolman:1934za,Bondi:1947av,
Romano:2010nc,Sinclair:2010sb,Marra:2010pg,Valkenburg:2011tm}.
The metric is given by 
\ba
\label{LTBmetric2}
\d s^2 = -\d t^2 + \frac{\apar^2(t,r)}{1-k(r)r^2}\d r^2 + \aperp^2(t,r)r^2\d\Omega^2 \,,\label{eq:metric}
\ea
where the radial ($\apar$) and angular ($\aperp$) scale factors are related by
$\apar = (\aperp r)'$. A prime denotes partial derivation with respect to the coordinate radius $r$.
The curvature $k=k(r)$ is 
a free function. The Friedmann-Lemaître-Robertson-Walker (FLRW) limit is $k\to$\,const., and $a_\perp=a_\|$.
The two scale factors define two Hubble rates:
\ba\label{H}
\Hperp= \Hperp(t,r) \equiv {\dot a_\perp}/{\aperp}\,,~~~~~~\Hpar=\Hpar(t,r) \equiv {\dot a_{\|}}/{\apar} \,.
\ea
The analogue of the Friedmann equation in this space-time is
then given by
$
H_{\perp}^2 = m(r) /a_{\perp}^3-k/a_{\perp}^2+ \Lambda / 3 \,,
$
where $m(r)$ is a non-negative free function of $r$ related to the locally measured matter density
$
8 \pi G \, \rho_{m}(t,r) = (m(r)  r^{3})' / a_{\parallel}a_{\perp}^2 r^2 \,,
$
which obeys the conservation equation
$
\dot{\rho_{m}}+ (2 H_{\perp}+H_{\parallel}) \rho_{m} =0 \,.
$
Dimensionless density parameters for the CDM and curvature are in analogy with the FLRW models:
\be \label{omegas}
\Omega_m(r)=\frac{m}{\Hperpo^2 } ,
~~~
\Omega_k(r)=-\frac{k}{\Hperpo^2},
~~~
\Omega_\Lambda(r)=\frac{\Lambda}{3\Hperpo^2 } ,
\ee
so that $\Omega_m(r)+ \Omega_k(r)+\Omega_\Lambda(r)=1$. Note that in the previous equation the gauge fixing $a_\perp(t_0,r)=1$ has been used.
Moreover, $\Omega_\Lambda$ depends on $r$ because the present-day expansion rate $\Hperpo$ is inhomogeneous.
Using (\ref{omegas}) the Friedmann equation takes on its familiar form:
$
{\Hperp^2}/{\Hperpo^2}=\Omega_m \, \aperp^{-3} + \Omega_k \, \aperp^{-2} + \Omega_\Lambda \,.
$
Integrating the Friedmann equation from the time of the big bang $t_{\rm bb}(r)$ to some later time $t$ yields the age of the universe at a given $(t,r)$:
$
t - t_{\rm bb} = \frac{1}{\Hperpo(r)}\int_{0}^{\aperp (t,r)} \!\!\!\!\!\!\!\!\! \frac{\d  x}{\sqrt{\Omega_m (r)x^{-1} + \Omega_k (r) +  \Omega_\Lambda(r) x^{2}}} .
$
Hence there is a relation between the functions $t_{\rm bb}$, $\Omega_k$ and $\Omega_m$. Therefore the $\Lambda$LTB model is specified by two free functional degrees of freedom, and we  use $\Omega_k(r)$ and $t_{\rm bb}(r)$.
By demanding a homogeneous age of the universe we fix the bang function to zero, $t_{\rm bb}(r)=0$.
This ensures the absence of decaying modes in the matter density~\citep{1977A&A....59...53S,Zibin:2008vj}, in agreement with the standard inflationary scenario.

We parametrize the only left freedom with the curvature function with the monotonic profile
\begin{align} \label{profi1}
k_{\alpha}(r)= k_{b} + (k_c - k_{b}) \; P_3 ( {r}/{r_b}, \alpha  ) \,,
\end{align}
where $r_b$ is the comoving radius of the spherical inhomogeneity and the function $P_{n}$ -- $C^n$ everywhere -- is:
\begin{align} \label{Pnf}
P_{n}(x,\alpha)= \left\{\begin{array}{ll}
1 & \mbox{ for }  0 \le x < \alpha \\
1-e^{- \frac{1-\alpha}{x-\alpha} \left(1-\frac{x-\alpha}{1-\alpha}\right)^n } & \mbox{ for }  \alpha  \le x < 1\\
0 & \mbox{ for } x \geq 1
\end{array}\right. . 
\end{align}
We choose $n=3$, such that the metric is $C^2$ and the Riemann curvature is $C^0$. Moreover, for $r \ge r_b$ the curvature profile equals $k_{b}$  such that there the metric describes {\em exactly} a curved $\Lambda$CDM model
 ($\alpha$ parametrizes the transition $k_c \rightarrow k_b$).
The central over- or under-density, determined by curvature $k_c$, is automatically compensated by a surrounding under- or over-dense shell.
Hence $r_{b}$, $k_{c}$ and $\alpha$ are the free parameters relative to the non-Copernican feature.

Finally, on the past light cone of a central observer,  $t(z)$ and  $r(z)$ are determined as a function of redshift $z$ by the differential equations for radial null geodesics,
$
\frac{\d t}{\d z} = \frac{-1}{(1+z)\Hpar}
$
and
$
\frac{\d r}{\d z} = \frac{\sqrt{1-k r^2}}{(1+z)\apar\Hpar} \,,
$
where $H_\|$ and $\apar$ are evaluated on the light cone. The area ($d_A$) and luminosity ($d_L$) distances are given by
$
d_A(z)=a_\perp \big(t(z),r(z) \big) \; r(z)$,   $d_L=(1+z)^2 d_A.
$

\section{Data \& Observables}

\noindent{\bf $\mathbf{H_0}$:}
The local Hubble rate is obtained by measuring cosmological standard candles mostly within a redshift range $z_{\rm min}\leq z \leq z_{\rm max}$ which depends on the redshift volume that is probed by a given experiment.
We compare the observed value to the theoretical quantity,
\begin{equation} \label{hloco}
H_{0}=  {c \, (z_{\rm max}-z_{\rm min})  \over \int_{z_{\rm min}}^{z_{\rm max}}  d_{L}(z)/[z+{1 \over 2}(1-q_{0})z^{2}]  \d z}  \,,
\end{equation}
where the deceleration parameter at the centre $q_{0}=\Omega_m (0)/2 - \Omega_\Lambda (0)$ is used to expand the luminosity distance to the second order in redshift.
The reason we compare an averaged expansion rate to the data is because the observed $H_0$ in fact comes from averaging distances, so this should be a fair comparison.
We use the value measured by \cite{Riess:2011yx} of $H_{\rm Riess}=73.8 \pm 2.4$  km\,s${}^{-1}$\,Mpc${}^{-1}$, with  $ z_{\rm min}=0.01$ and $ z_{\rm max}=0.1$.
One could be tempted to say that modern cosmic-microwave-background (CMB) data constrain H$_0$ already sufficiently without the inclusion of the astrophysical data that we discuss here. However, it is crucial to realise that {\em only in a purely homogeneous universe}, these two measurements measure the same quantity. When considering the non-Copernican feature at hand, the expansion rate varies spatially and hence a model can predict different values for the local astrophysical H$_0$ (local expansion rate) and the CMB H$_0$ (the age of the universe and inferred from it the global expansion rate today).

\noindent{\bf Supernovae Ia:}
We use the SNLS3 catalogue~\citep{Guy:2010bc}, which consists of 472 type Ia supernovae in the redshift range $z=0.01-1.39$. We include two nuisance parameters describing stretch-luminosity and colour-luminosity relationships, as in~\cite{Guy:2010bc}.

\noindent{\bf CMB:}
We fit the CMB according to the method presented in~\cite{Moss:2010jx,Biswas:2010xm}, in which an effective FLRW metric is used to account for the different area distance to the surface of last scattering as compared to the homogeneous background model.
This method ignores isocurvature modes (consistent with the choice of a homogeneous big bang), assumes a standard number of relativistic degrees of freedom and a standard power spectrum, all of which would change the constraints~\citep{Clarkson:2012bg}. Moreover, we assume that the late-time integrated Sachs-Wolfe effect is not affected by the presence of the inhomogeneity, as the latter will turn out to be of a magnitude such that it can be described as linear
perturbations on an FLRW metric (see the discussion below about the kSZ observable).
We fit our model to WMAP 7-year data~\citep{Komatsu:2010fb}.

\noindent{\bf Baryon Acoustic Oscillations (BAO):}
The sound horizon at the time of the drag epoch $t_d$ is imprinted in the galaxy correlation function.
In a spherically symmetric inhomogeneous model this is an ellipsoid with proper scales, when viewed from the centre 
\begin{align}
L_\perp(z) = & \, d_s \; {a_\perp(z)}/{a_\perp(t_d,r(z))}=d_A(z) \, \theta_{s}(z) \,,  \\
L_\|(z) =& \, d_s \; {a_\|(z)}/{a_\|(t_d,r(z))}={z_{s}(z)}/{[(1+z)H_\|(z)]} \,,
\end{align}
where 
$\theta_{s}$ is the angle that the acoustic scale subtends on the sky, and $z_{s}$ is the redshift interval corresponding to the acoustic scale in the radial direction.
The sound horizon $d_s$ is calculated assuming a homogeneous early universe. We use observations from SDSS, 6DFGS and WiggleZ~\citep{Percival:2009xn,Beutler:2011hx,Blake:2011en}, as compiled in~\cite{Zumalacarregui:2012pq}.

\noindent{\bf Compton $\mathbf{y}$-distortion:}
Off-centre observers see a large dipole in the CMB in an inhomogeneous universe. CMB photons are scattered from inside our past light-cone into our line-of-sight by off-centre reionized structures which act as mirrors.
The spectrum observed by the central observer is then a mixture of blackbody spectra with different temperatures, producing a distorted blackbody spectrum.
In the single-scattering and linear approximations, and when the temperature anisotropy is dominated by the induced dipole $\beta$, the $y$-distortion can be written as~\citep{Moss:2010jx}:
\begin{equation} \label{ydist}
y = \frac{7}{10}\int_0^{r_{\rm re}} \d r \, \frac{\d\tau}{\d r} \, \beta(r)^2 \,,
\end{equation}
where $r$ is the comoving distance down the light cone and $r_{\rm re}$ marks the reionization epoch. The time dependence of the optical depth is given by
$
\d\tau / \d t = \sigma_T \, n_{e}(t)= \sigma_T \,  f_{\rm b} \, (1 - Y_{\rm He}/2) \, \rho_{\rm m}(t) /m_p \,,
$
where $\sigma_T$ is the Thomson cross-section, $f_{\rm b} \equiv 
\rho_{\rm b}/\rho_{\rm m}$ is the baryon fraction, $Y_{\rm He}$ is the 
helium mass fraction and $m_p$ is the proton mass.
The dipole $\beta$ is found by integrating the geodesic equations 
in the negative and positive $r$-directions starting from an observer at $\{t(z),r(z)\}$ back to the surface of last scattering.
The difference in redshift between the two directions is then approximately translated into the dipole observed by the scatterer:
$
\beta=  (z_{+}-z_{-}) / (2+z_{ +}+z_{-}) \,.
$
The $2\sigma$ upper limit from the COBE satellite~\citep{Fixsen:1996nj} is $y < 1.5\times10^{-5}$.

\noindent{\bf kSZ:}
The dipole $\beta$ affects the observed CMB also through the kinetic Sunyaev-Zel'dovich (kSZ) effect~\citep{GarciaBellido:2008gd}: hot electrons inside an overdensity distort the CMB spectrum through inverse Compton scattering, in which low energy CMB photons receive energy boosts during collisions with the high-energy electrons.
Here we focus on the `linear kSZ effect' \citep{Zhang:2010fa}, in which the effect due to all free electrons in the reionized universe is taken into account.
Using the Limber approximation, the kSZ power at multipole $\ell$ is given by~\citep{Moss:2011ze}:
$
C^{{\rm kSZ}}_\ell \simeq \frac{16\pi^2}{(2\ell + 1)^3}
          \int_0^{r_{\rm re}} \! \d r \, r \! \left[ \beta(r)  \frac{\d \tau}{\d r} \right]^{2}  \!\!  \Delta_{m}^{2}  \left( \hat k (r) ,r\right)
$,
where $\Delta_{m}^{2}(k,z) ={k^{3} \over 2 \pi^{2}} P_{m}(k,z)$ is the dimensionless power spectrum of the background model
and the function $\hat k(r)\equiv \hat k(k(r),z(r))$ is necessary to ``isotropize'' the angular and radial wave numbers which in an inhomogeneous universe evolve differently:
$
\hat k(\bar k,z) = \bar k  [ (1+ \bar z) a_\perp(\bar t,r(z))^{2/3} a_\|(\bar t,r(z))^{1/3} ] /[(1+z)  a_\perp(z)^{2/3} a_\|(z)^{1/3}] .
$
We constrain inhomogeneous models using a top hat prior $0<l(l+1)\left[C^{\rm TT}_{\ell=3000}+C^{\rm kSZ}_{\ell=3000}\right]<59\mu{\rm K}^2$, based on the results from SPT~\citep{Shirokoff:2010cs}.

As we will see in the results, the use of a matter power spectrum that is computed in the FLRW metric, $\Delta_{m}^{2}(k,z)$, even though the correct metric is the LTB metric, is justified. Density contrasts that we encounter in the LTB metric are of magnitudes such that they could be described as a linear perturbation on the FLRW metric. Therefore, any deviation of $\Delta_{m}^{2}(k,z)$ from its FLRW evolution as a consequence of the spherical perturbation, is of second order in perturbation theory and hence suppressed since both $\Delta_{m}^{2}(k,z)$ and the spherical perturbation are at the linear level. This argument may sound contradictory to the findings of~\cite{Nishikawa:2012we}. However, in~\cite{Nishikawa:2012we} it is found that structures with a central density $\delta_0=-1$ and radius $L=12$Gpc have a radius-dependent growth factor significantly different from FLRW. Such structures can clearly not be considered a linear perturbation ($|\delta|\sim1$)
and one cannot expect that superposed perturbations continue to behave uncoupled (linearly). In the scenario at hand, the assumption that perturbations remain linear, and hence that an FLRW $P(k)$ superposed on a linear LTB perturbation grows effectively as in pure FLRW, is justified.

\noindent{\bf Age data:}
Finally, we constrain radial inhomogeneity also by means of galaxy ages~\citep{Bolejko:2011ys,McCarthy:2004hx,Simon:2004tf,Stern:2009ep,Moresco:2012jh,Wang:2011kj,dePutter:2012zx}. In particular, we use the cosmology independent age vs.\ redshift data as compiled in~\cite{dePutter:2012zx}, which in the most conservative approach only provides a lower bound on cosmic ages. For completeness we present results using these ages as lower bounds as well as using them as absolute age measurements (see~\citealt{dePutter:2012zx} for details).

\begin{figure}
\begin{center}
\includegraphics[width=0.85\columnwidth]{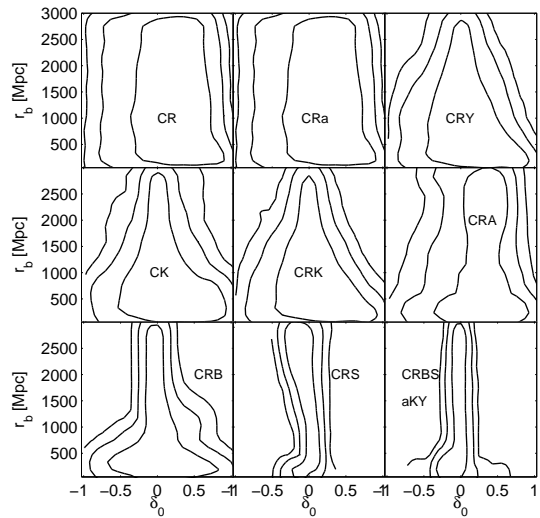}
\caption{Marginalized constraints on $r_{b}$ and $\delta_{0}$ from different combinations of data: ``C'' refers to CMB, ``R'' to $H_0$, ``B'' to BAO, ``S'' to SN, ``A'' to absolute age data, ``a'' to age data used as lower bound only, ``K'' to kSZ, ``Y'' to Compton-y distortion.
The strongest constraints at low redshifts come from SN, while the strongest constraint on $\delta_{0}$ on large scales seems to come from kSZ. Naturally all data sets combined give the strongest constraints.
}\label{fig:alldata}
\label{default}
\end{center}
\end{figure}

\section{{Copernican prior}} \label{sec:copprior}
Given a Gaussian density field, the 
mean square of density perturbations inside a sphere of radius $L$ around any point (hence also around the observer) today is given by
$\sigma_L^{2} = \int_0^\infty \frac{\d k}{k}\Delta_{m0}^{2}(k)  \left[ 3j_1 (Lk)/Lk\right]^2$,
where $\Delta_{m0}$ is the power spectrum today inferred from the CMB temperature spectrum, assuming a Copernican universe, and $j_l$ is the spherical Bessel function of the first kind. 
We calculate $\sigma_L$ for the radius $L<r_{b}$ at which the central over/under-density makes the transition to the surrounding mass-compensating under-/over-dense shell.
Then we compute the actual density perturbation $\delta_0 \equiv M(L) / \bar M(L) -1$ of a given inhomogeneity and define the Copernican prior as the probability
\begin{equation}
P(\delta_0,L) = (\sigma_{L} \sqrt{2\pi})^{-1} \exp \left[ -\tfrac{1}{2} \left(\delta_0/\sigma_L\right)^2\right],
\end{equation}
where $M(r)\equiv 4\pi \int_0^r dr \sqrt{-g} \rho_m(r)$ is the mass of the inhomogeneity, $\bar M (r)= \left. M(r)\right|_{k(r)=k_{b}}$ is the mass relative to the background, and $g$ the determinant of the metric~\eqref{eq:metric}. A similar function has been used by~\cite{Hunt:2008wp} to compute the probability of having a large void in an Einstein-de Sitter universe, so large that the need for a cosmological constant vanishes (which is not the case considered here).

\section{{Analysis}} 
We explore a 14-dimensional parameter space using {\sc CosmoMC}~\citep{Lewis:2002ah}: three parameters describing the inhomogeneity, 
an overall curvature term,
the ratio of baryons to dark matter, the cosmological constant, the optical depth to the last scattering surface, the age of the universe, three spectral parameters for the CMB (scalar amplitude, tilt and running of the tilt), the amplitude of a thermal SZ template that is used for the CMB spectrum and the two SN nuisance parameters.
We calculate cosmic distances using {\sc VoidDistancesII}~\citep{Marra:2012pj}, and compute the corrected CMB spectrum using {\sc CAMB}~\citep{Lewis:1999bs}.

\begin{figure}
\begin{center}
\includegraphics[width=0.9\columnwidth]{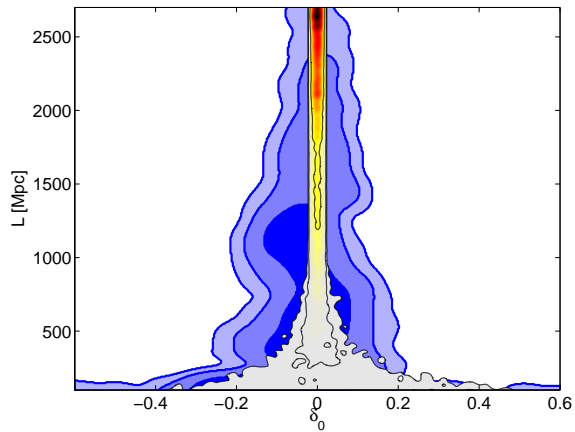}
\caption{Marginalized posterior probability of $L$ and $\delta_{0}$ from all data sets combined (blue contours, closely related to CRBSaKY (CMB, H$_0$, BAO, SN, age data, kSZ and Compton-y) in Fig.~\ref{fig:alldata}, albeit with a slightly different quantity on the vertical axis)
at 68\%, 95\% and 99\% confidence level (c.l.), compared to the Copernican posterior obtained by fitting CMB only while imposing the Copernican prior, (red to gray colouring) at 68\%, 95\% and 99\% c.l. The ratio of the areas of the 99\%-c.l.\ surfaces is roughly a factor of three.
The allowed area needs to be reduced by a factor of three so as to confirm the Copernican principle.
The marginalized 95\% c.l.\ limits on the contrast are $-0.29<\delta_{0}<0.14$ for CRBSaKY and $-0.12<\delta_{0}<0.12$ for the Copernican prior.}
\label{fig:awesome}
\end{center}
\end{figure}

Given the curvature parameter $k_{c}$ that describes the spherical patch, in Eq.~\eqref{profi1}, we compute the density contrast $\delta_0$ as the ratio of the determinant of the metric~\eqref{eq:metric} at $r=0$ and $r>r_b$, since the coordinates are synchronous and comoving. In fact, we take a flat prior on the central density contrast $-1<\delta_0<1$, and obtain the actual curvature parameters in each step of the Monte Carlo Markov Chain (MCMC) by numerical inversion. Fundamentally, $\delta_0$ is mathematically not constrained from above, such that our prior looks artificially constraining. As we will see however, the observations favour $\delta_0<1$, such that we can safely state that our result is prior independent. Also, we take a flat prior on $r_{b}$ and $\alpha$ on the ranges $50 \text{Mpc}\! <\! r_{b}\! <\! 3000$Mpc and $0\! <\! \alpha \! <\! 1$. 
There is a slight degeneracy between $r_b$ and $\alpha$, as a smoother transition (smaller $\alpha$) and larger radius 
give the same difference in gravitational potential at the centre and outside of the feature. However -- as can be seen by comparing Fig.~\ref{fig:awesome} to the lower right panel in Fig.~\ref{fig:alldata} -- constraints on
$L$
do not differ qualitatively from constraints on $r_b$, while the relation between $r_b$ and $L$ depends on $\alpha$. This implies that the degeneracy does not significantly bias the results.

We will consider all previously discussed observables in conjunction with CMB and H$_0$ data. The reason for this is that in all cases we use the same set of 11 ($\Lambda$CDM) + 3 (non-Copernican feature) parameters and we wish not to bias the constraining power of any data set by fixing any parameter that at first sight might seem unrelated. At the same time, none of the individual data sets can constrain all 14 parameters all by itself, which explains the need for the inclusion of the CMB and H$_0$. Moreover, the kSZ and Compton-y effects depend directly on the CMB data, such that it would not make sense to consider them without computing the CMB power spectrum. The kSZ effect is nothing but an alteration of the CMB power spectrum, while the Compton-y effect depends on the spectral distortions, which do not enter the power spectrum. That is, no data is double counted. In summary, we do not want to fix any parameter, but having many parameters completely unconstrained is inconvenient for the MCMC analysis and hence we always include CMB and H$_0$.

\section{{Results}}

\begin{figure}
\begin{center}
\includegraphics[width= 0.75\columnwidth]{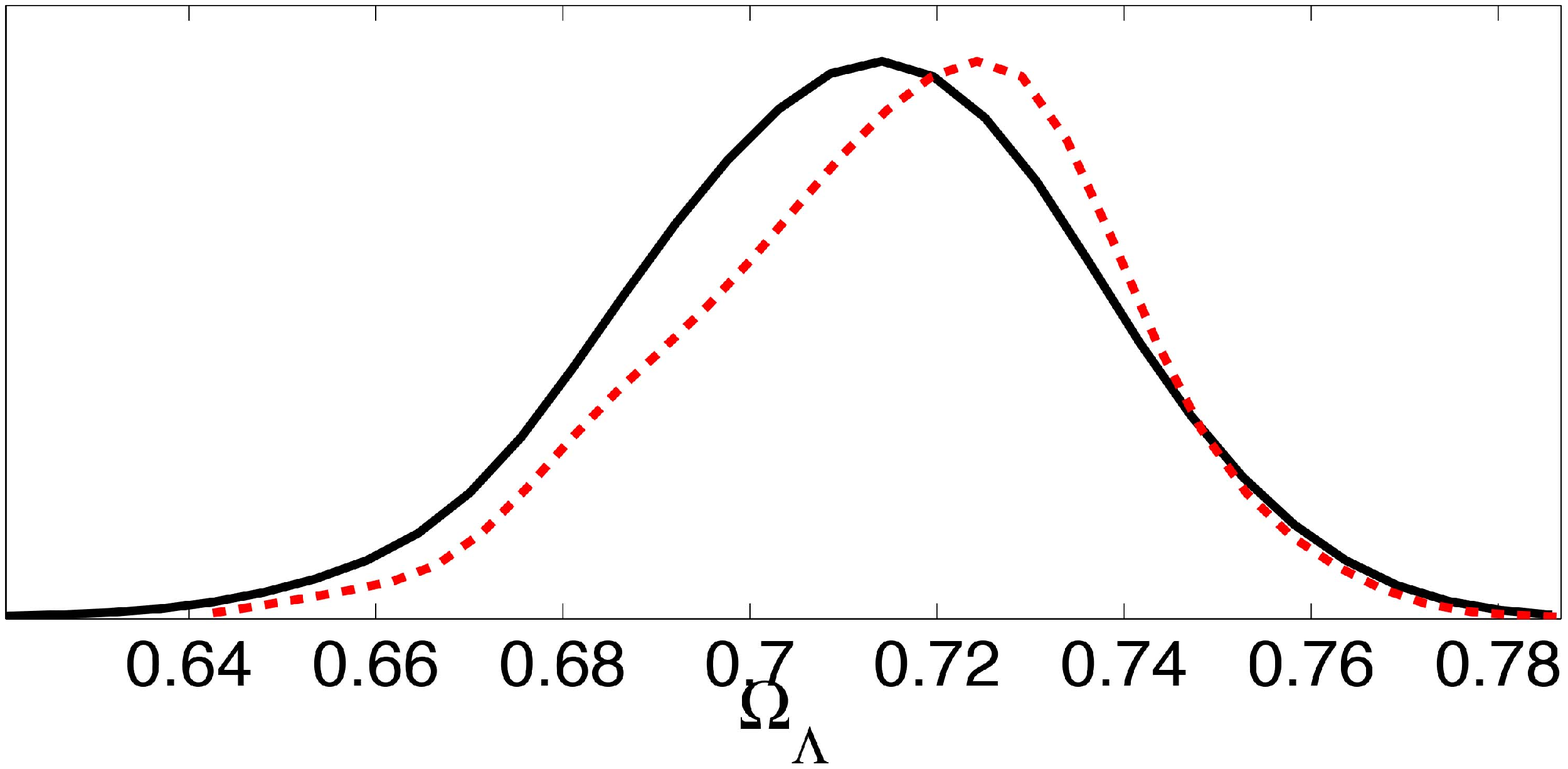}
\caption{Marginalized posterior probability for $\Omega_\Lambda$
when ignoring inhomogeneity (red dashed line), $\Omega_\Lambda=0.71\pm0.05$ at 95\% c.l., and when including inhomogeneity and marginalizing over it (solid black line), $\Omega_\Lambda=0.76\pm0.05$ at 95\% c.l. Both constraints correspond to the CRBSaKY observables, see Fig.~\ref{fig:alldata}.
}
\label{fig:omega_l}
\end{center}
\end{figure}

We present our results in terms of the non-Copernican parameters $\delta_0$ and $r_b$, marginalizing over all other aforementioned parameters. 
In Fig.~\ref{fig:alldata} we show the constraints on the non-Copernican parameters from a number of possible combinations of data sets, ordered by constraining power. In all cases we use the CMB constraints in combination with at least one other data set. 
Not surprisingly, all data sets combined provide the strongest constraints, only allowing for a narrow range of contrasts, however for many radii.
As warned for in the introduction, none of the observables actually favors $\delta_0 \neq 0$. But as argued, the purpose of this Letter is not to establish the Copernican principle,
but instead to falsify the extra parameters observationally.

The first panel shows that the combination of CMB and H$_0$ data is hardly constraining on the radius of the non-Copernican feature, and only mildly on the central density. This can be understood by realizing that the central expansion rate is linearly related to the central density contrast, and not at all to the radius, while the CMB is only a standard candle at very high redshift, which does almost not depend on the non-Copernican feature.
The third, fourth and fifth panels isolate the constraining power from the kSZ effect and Compton-y distortion. Both these effects depend on the anisotropy of the CMB at outer radii, which is to first approximation sensitive to the depth of the gravitational potential of the non-Copernican feature. Keeping the density contrast constant, but increasing the radius, will increase the depth of the potential. Indeed, these observables constrain mostly the larger radii. 
The age data 
used as a lower bound hardly constrain the inhomogeneity, while the absolute data
which have both an upper and a lower bound merely serve as an illustration, as these cannot be used in the LTB metric~\citep{dePutter:2012zx}. At last, the observables CMB, H$_0$, BAO and SNe are a set that for this analysis is an angular diameter distance measurement at many redshifts. As such, the SNe are most strongly constraining at small radii, because a large value of $\delta_0$ at these radii cannot transit back to zero sufficiently quickly without predicting a bump in the luminosity distance 
that violates SN observations.

The key result of this Letter is shown in Fig.~\ref{fig:awesome}, where we compare the posterior probability from all data sets to a posterior which is the Copernican prior convolved with the CMB likelihood,
$
\tilde P(\delta_0,L)  =
\int dp_i P(\delta_0,L) \mathcal{L}_{\rm CMB}(p_i,\delta_0,L)
$,
where $p_i$ denote {\em all} parameters other than $\delta_0$ and $L$, and $\mathcal{L}_{\rm CMB}(p_i,\delta_0,L)$ is the likelihood of the CMB data given the model and its parameters $p_i$, $\delta_0$ and $L$. In words, 
$\tilde P$ is the probability distribution of $\delta_0$ and $L$, given the matter power spectrum obtained from the CMB and its uncertainty, which under the Copernican principle is the power spectrum around us.
We find that the area of constraints on $L$ and $\delta_0$ needs to decrease by a factor of three to confirm the Copernican principle, because then observations on the lightcone constrain inhomogeneity to within the range allowed for by the CMB power spectrum. 
Finally, in Fig.~\ref{fig:omega_l} we show for the first time constraints on the cosmological constant, $\Lambda$, marginalized over the effect of inhomogeneities around us, compared to the same constraints without taking into account inhomogeneity. We find that error bars increase by 15\% if one marginalizes over inhomogeneity.

The analysis regarding the constraints from $H_{0}$, SNe, BAO and kSZ uses data and results that have been obtained assuming, at some point, an FLRW framework. While we correctly used the processed data so as to compare it with the inhomogeneous universe, in principle one should confront to data that are as close to raw as possible. While this caveat should be kept in mind, 
we do not expect the analysis to be sizeably biased because the inhomogeneous universe we are considering does not depart strongly from its FLRW background, as shown by the results of Fig.~\ref{fig:awesome}.
Our constraints apply to inhomogeneity which arise at late times, and do not apply to 
non-Copernican properties at early times.
We assumed that dark energy is described by a cosmological constant and that general relativity is the correct theory of gravity.

\section*{Acknowledgements}
WV is supported by a Veni research grant from the Netherlands Organisation for Scientific Research (NWO).

\bibliographystyle{mn2e}
\bibliography{refs.bib}

\label{lastpage}

\end{document}